\documentclass[aps,prl,twocolumn,
superscriptaddress,reprint]{revtex4-1}
\usepackage{graphicx}
\usepackage{setspace}
\usepackage{amsmath}
\usepackage{amssymb}
\usepackage{color}
\usepackage{array}
\usepackage{subfigure}
\usepackage{hyperref}
\usepackage{float}
\usepackage{lipsum}
\usepackage{ulem}

\usepackage[all]{xy}
\newcommand{\RN}[1]{%
  \textup{\uppercase\expandafter{\romannumeral#1}}%
}

\begin{document}
\title{Digital Quantum Simulation of Hadronization in Yang-Mills Theory}
\author{De-Sheng Li}
\affiliation{Interdisciplinary Center for Quantum Information, National University of Defense Technology, Changsha 410073, P.R.China}
\affiliation{College of Physics, Mechanical and Electrical Engineering, Jishou University, Jishou 416000, P.R.China}

\author{Chun-Wang Wu}
\affiliation{Interdisciplinary Center for Quantum Information, National University of Defense Technology, Changsha 410073, P.R.China}

\author{Ming Zhong}
\email[email: ]{zhongm@nudt.edu.cn (corresponding author)}
\affiliation{Department of Physics, National University of Defense Technology, Changsha 410073, P.R.China}

\author{Wei Wu}
\affiliation{Interdisciplinary Center for Quantum Information, National University of Defense Technology, Changsha 410073, P.R.China}

\author{Ping-Xing Chen}
\email[email: ]{pxchen@nudt.edu.cn}
\affiliation{Interdisciplinary Center for Quantum Information, National University of Defense Technology, Changsha 410073, P.R.China}

\begin{abstract}
A quantum algorithm of $SU(N)$ Yang-Mills theory is formulated in terms of quantum circuits. It can nonperturbatively calculate the Dyson series and scattering amplitudes with polynomial complexity. The gauge fields in the interaction picture are discretized on the same footing with the lattice fermions in momentum space to avoid the fermion doubling and the gauge symmetry breaking problems. Applying the algorithm to the quantum simulation of quantum chromodynamics, the quark and gluon's wave functions evolved from the initial states by the interactions can be observed and the information from wave functions can be extracted at any discrete time. This may help us understand the natures of the hadronization which has been an outstanding question of significant implication on high energy phenomenological studies.
\end{abstract}
\maketitle


\section{Introduction}
\label{intro}
Yang-Mills theory plays a fundamental role in the constructions of the Standard Model (SM) of particle physics. One important property of the theory is asymptotic freedom demonstrating that the interaction becomes weaker as the energy scale evolves higher. This makes the theory rigidly predictable by perturbation when the energy of the physics processes are high enough. While in the opposite infrared limit, the interaction is so large that the perturbation theory breaks down and some interesting phenomena are {very difficult to understand}. One significant example of the non-perturbative effects is the colour confinement in quantum chromodynamics (QCD) which is a prototype of Yang-Mills theory with $SU(3)$ colour symmetry. Due to the confinement, the colour charged quarks and gluons can not be isolated and observed directly at low energy. They must clump together to form observable hadrons. 

The nature of hadronization is far away from well explored and remains as a big mystery in high energy physics. On the other hand, the knowledge of hadronization is indispensable in investigating the new physics beyond the SM, since hadrons appear inevitably in every high energy physics experiment. Plenty of endeavours have been made since 1960s and various phenomenological models can give us some information on hadronization so that it can be taken into account in the experimental data analysis. More precise and comprehensive knowledge of the hadronization effects is required as more precise experiments devoting to study the new physics. 

The most promising attempt to account for the effects in the near future is believed to employ the lattice theory \cite{wilson1974confinement}, a method to quantize a field theory on a  discretized lattice and permit numerical calculations in strong-coupling limit. Though it has achieved much progress, the capability of the lattice computations relies heavily on the performance of computers available, which in turn hinders it's applications since the development of the classical computer meets the bottleneck increasingly.

The breakthrough of the bottleneck is the quantum computer introduced notionally by Feynman three decades ago \cite{feynman1982simulating}, and has speeded up development recently \cite{Lloyd1996Universal,Lanyon2011Universal,Georgescu2014Quantum,Lamata2014efficient,Paraoanu2014recent,barends2015digital,somaroo1999quantum}.  The quantum computer is structurally powerful to simplify the calculation and adaptable to the complex system studies in essence \cite{Cirac2012goals,Blatt2012quantum,Arrazola2016digital-analog}. The idea of using it to explore the physics system of quantum field theory has set out to attract increasing interest. Specific scenarios, in view of trapped ions and ultracold atoms etal, were proposed to study various field theory models in four and fewer dimensions of spacetime \cite{Casanova2011Quantum,Xiang2018Experimental,notarnicola2015discrete,hauke2013quantum,marcos2013superconducting,rico2014tensor,zohar2017digital}. The quantum algorithms to calculate the scattering amplitudes of scalar quartic self-interaction and fermionic quantum field theory were developed and found to have exponential speedup \cite{Jordan2011Quantum, Jordan2014Quantum,jordan2014quantum2, moosavian2018faster, klco2019digitization}. The experimental simulation of Schwinger model in trapped ion system using 20 ions was implemented \cite{martinez2016realtime,Kokail:2018eiw}. For Yang-Mills theory, some theoretical proposals to use trapped ions, ultracold atoms, optical lattices and quantum computer to simulate were presented in \cite{byrnes2006simulating,zohar2013cold-atom,banerjee2013atomic,tagliacozzo2013simulation,wiese2014towards,zohar2016quantum,bender2018digital,Mezzacapo:2015bra,zohar2015formulation, dalmonte2016lattice, banuls2019review}. One impressive common feature of these studies is that quantum simulation works in both perturbative and non-perturbative regimes.

Given it's significance in high energy physics, it is deserved to make further efforts in universal quantum simulation of Yang-Mills theory. In this paper, we present a scheme to describe the $SU(N)$ Yang-Mills theory by digital quantum simulator and work out a quantum algorithm. As compared to the existed works which were treated in Schr$\ddot{o}$dinger picture and spacetime discretization, we apply the momentum space discretization in interaction picture. The advantages brough about are clear and competitive, since the interaction picture and momentum representation are the ways quantum field theories being actually applied to high energy physics. The algorithm is rigorous in design since the problem of fermions doubling can be avoided naturally and the spins of fermions and gauge bosons can be strictly handed. In application, it is more convenient when being exerted on simulating the processes in colliders. 

We begin with a short introduction to the formalism of $SU(N)$ Yang-Mills theory and it's canonical quantization in Sec. \ref{canqua}. The quantization to Yang-Mills gauge theory is rather different and complicated as compared to the pure scalar and abelian gauge theory, in that ghost fields are introduced into the Lagrangian accompanying by the gauge fixing. With these indispensable knowledge, we present an elaboration on the quantum simulation of the theory in Sec. \ref{quasim}. We formulate a canonical quantization based algorithm to calculate the Dyson series of the theory non-perturbatively and thus can be used to study the system in both perturbative and non-perturbative regimes. Using qubits the number of which increases polynomially with the lattices to simulate the interactions, the algorithm has exponential speedup. Moreover, the dynamical information from the evolving wave functions of particles can be extracted at any discrete time. In Sec. \ref{application}, we discuss the possible application on studying the hadronization of QCD. A short summary and comment is made in the last section, where an estimate on the amounts of quantum devices for a minimum simulation is made. We stress that the algorithm is not practical at moment for the unreachable required amount of quantum devices though it may help us understand the low energy QCD in the future when quantum computer is powerful enough.

\section{Canonical Quantization of Yang-Mills Theory}
\label{canqua}

For definiteness and practicality, we study a one-flavor non-abelian  gauge theory with $SU(N)$ symmetry. The extension to arbitrary $N_f$ duplicates of fermions is straightforward. The Lagrangian density is \cite{Kugo1978Manifestly}
\begin{eqnarray}\label{L}
\mathcal{L}&=&-\frac{1}{4}(F^{a}_{\mu\nu})^{2}+\bar{\psi}_{i}(i\delta_{ij}\gamma^{\mu}\partial_{\mu}+g\gamma^{\mu}A_{\mu}^{a}t^{a}_{ij}-m\delta_{ij})\psi_{j}\nonumber\\  
&&-\frac{1}{2}(\partial^{\mu}A_{\mu}^{a})^{2}+(\partial^{\mu}\bar{c}^{a})\partial_{\mu}c^{a}+g f^{abc}(\partial^{\mu} \bar{c}^{a})A^{b}_{\mu}c^{c},
\end{eqnarray}
where $\psi_{i}$ and $A^{a}_{\mu}$ are fermion and gauge fields with $i=1,\cdots,N$ and $a=1,\cdots,N^2-1$. The Dirac matrices $\gamma^{\mu}$ are defined so that $\{\gamma^{\mu},\gamma^{\nu}\}=2\eta^{\mu\nu}I_{4\times 4}$ with the convention $\eta^{\mu\nu}=diag(1,-1,-1,-1)$ for the spacetime metric. The $SU(N)$ generators $t^{a}$ are represented by the traceless and Hermitian matrices. The gauge field strength reads
\begin{eqnarray}
F^{a}_{\mu\nu}=\partial_{\mu}A^{a}_{\nu}-\partial_{\nu}A^{a}_{\mu}+gf^{abc}A^{b}_{\mu}A^{c}_{\nu},
\end{eqnarray}
where $f^{abc}$ are the structure constants of the group. The first two terms in Eq.~(\ref{L}) compose the Yang-Mills theory. To make a proper quantization, one must introduce the gauge fixing. We adopt the Feynman gauge which is Lorentz invariant and this results in Faddeev-Popov ghosts in the other terms. The longitudinal gauge fields growing out of the gauge fixing and the ghost fields $c^{a} ( \bar{c}^{a})$ do not correspond to physical particles at all. Since the ghosts are Grassmann fields, there appears a minus sign when adjacent ghost field being swapped. The Hermiticity condition of Lagrangian density yields to \cite{Kugo1978Manifestly}
\begin{eqnarray}
\bar{c}^{a\dagger}(x)=-\bar{c}^{a}(x),&& c^{a\dagger}(x)=c^{a}(x).
\end{eqnarray}

We canonically quantize the  theory in interaction picture where the fields behave freely so that the field operators are expanded in terms of the orthonormal and complete solutions of the free equations of motion with the annihilation and creation operators as the coefficients
\begin{eqnarray} \label{qfields}
A^{a}_{\mu}(x)\!&=&\!\int \frac{d^{3}p}{(2\pi)^{3}}\frac{1}{\sqrt{2\omega_{\vec{p}}}}\sum \limits_{l}(a^{a}_{\vec{p},l}\epsilon^l_{\mu}e^{-ip\cdot x}+a^{a\dagger}_{\vec{p},l}\epsilon^{l*}_{\mu}e^{i p\cdot x}), \nonumber\\
\psi_{i}(x)\!&=&\!\int \frac{d^{3}p}{(2\pi)^{3}}\frac{1}{\sqrt{2\omega_{\vec{p}}}}\sum \limits_{s}(b_{\vec{p},i}^{s}u^{s}_{p}e^{-ip\cdot x}\!+\!c^{s\dagger}_{\vec{p},i}v^{s}_{p}e^{ip\cdot x}), \nonumber\\
{c}^{a}(x)\!&=&\!\int \frac{d^{3}p}{(2\pi)^{3}}\frac{1}{\sqrt{2\omega_{\vec{p}}}}(d_{\vec{p}}^{a}e^{-ip\cdot x}+d^{a\dagger}_{\vec{p}}e^{ip\cdot x}),\nonumber\\
\bar{c}^{a}(x)\!&=&\!\int \frac{d^{3}p}{(2\pi)^{3}}\frac{1}{\sqrt{2\omega_{\vec{p}}}}(e_{\vec{p}}^{a}e^{-ip\cdot x}-e^{a\dagger}_{\vec{p}}e^{ip\cdot x}).
\end{eqnarray}
The $\omega_{\vec{p}}=p^{0}=\sqrt{|\vec{p}|^{2}+m^{2}}$ denotes the energy of the particles created or annihilated by  the coefficient operators. The $\epsilon^{l}_{\mu}$ are polarization vectors of the massless gauge fields. The $u^{s}_{p}$ and $v^{s}_{p}$ are the Dirac spinors for the particles and anti-particles respectively. The following equations are used to ghost field in canonical quantization
\begin{eqnarray}
\frac{\partial (\dot{\bar{c}}^{a}\dot{c}^{a})}{\partial \dot{c}^{a}}=\dot{\bar{c}}^{a},&&\frac{\partial (\dot{\bar{c}}^{a}\dot{c}^{a})}{\partial \dot{\bar{c}}^{a}}=-\dot{c}^{a},
\end{eqnarray}
where $dot$ upon ghost field means the partial time derivative.

Giving the equal time commutation to gauge field and anti-commutation relations to the other fields with their canonical momentums, we then have the relations in terms of the annihilation and creation operators
\begin{eqnarray}
[a^{a}_{\vec{p},l},a^{b\dagger}_{\vec{p}^{\prime},m}]\!&=&\!(2\pi)^{3}\delta^{(3)}(\vec{p}-\vec{p}^{\prime})\delta_{lm}\delta^{ab},\nonumber\\
\{b_{\vec{p},i}^{r}, b^{s\dagger}_{\vec{p}^{\prime},j} \}\!=\!\{c_{\vec{p},i}^{r}, c^{s\dagger}_{\vec{p}^{\prime},j}\}\!&=&\!(2\pi)^{3}\delta^{(3)}(\vec{p}-\vec{p}^{\prime})\delta^{rs}\delta_{ij},\nonumber\\
\{d^{a}_{\vec{p}}, e^{b\dagger}_{\vec{p}^{\prime}} \}\!=\!\{e^{b}_{\vec{p}},d^{a\dagger}_{\vec{p}^{\prime}}\}\!&=&\!-(2\pi)^{3}\delta^{(3)}(\vec{p}-\vec{p}^{\prime})\delta^{ab},
\end{eqnarray}
with all the other commutators or anticommutators vanished. Note that we have the anti-commutation relation for the annihilation operator and the creation operator of ghost and anti-ghost field in the last equation.

\section{Quantum Simulation of Yang-Mills Theory}
\label{quasim}

We now switch to the quantum simulation. The main tasks involve giving maps between qubits and particle states, digitizing the interactions by matrices and evolving the qubits by the operations of the matrices via quantum circuits. We study within the interaction picture where only the interaction terms are involved to the representation in terms of quantum circuits. The simulation in the interaction picture can lead to an exponential improvement in gate complexity with an arbitrarily small error and substantially improve the efficiency for simulating some classes of diagonally dominant Hamiltonian \cite{low2018hamiltonian}.
 
To avoid the fermion doubling problem, the discretization in momentum space instead of the position space is performed. Attempts to implement such purpose have been carried out to the lattice fermions with the gauge field defined on links joining lattice points \cite{drell1976variational,Drell:1976mj}. Applying this formalism to the weak-coupling perturbation theories, it was found to suffer from the problems of Lorentz non-covariance and non-locality \cite{karsten1978,karsten1979the,karsten1981,Rabin:1981nm}. When discretizing the gauge fields on the same footing with lattice fermions in the momentum space, it was found that, in addition to absence of fermion doubling, the gauge symmetry is reserved  \cite{berube1990gauge,berube1991yang-mills}. We will use this non-compact formulation in discretization.

\subsection{Qubits representation of Yang-Mills theory}
The particles are described by the Fock states labeling with the momentum, spin and color charge as the degrees of freedom. To have a digital quantum simulation, the continuous three dimensional momentum space needs to be discretized to the lattice space 
\begin{eqnarray}
\Gamma=  \kappa\mathbb{Z}^{3}_{\hat{P}},
\end{eqnarray}
where the $\mathbb{Z}^{3}_{\hat{P}}$ denotes a $\hat{P}\times \hat{P}\times \hat{P}$ lattice in three dimensional momentum space. The $\hat{P}$ is equal to $Int(P/\kappa)$ with $P$ the range of momentum component and $a$ the lattice spacing.  The number of the lattice sites is $\mathcal{V}=\hat{P}^3$.

The multi-particle state of quantum field theory is Fock space valued. The bases to expand the Fock space of the Yang-Mills theory are
\begin{eqnarray}\label{F}
F=\{\otimes_{\vec{p}\in \Gamma} |n^{a}_{l},m_{i,\triangle}^{s},m_{j,\nabla}^{r},l^{b}_{\triangle},l^{c}_{\nabla}\rangle _{\vec{p}} \},
\end{eqnarray}
where $n^{a}_{l}$, $m_{i,\triangle}^{s}$, $m_{j,\nabla}^{r}$, $l^{b}_{\triangle}$, $l^{c}_{\nabla}$ are occupation numbers of gauge boson, fermion, anti-fermion, ghost and anti-ghost at lattice site $\vec{p}\in \Gamma$ respectively. They are all integers with $0\leq n^{a}_{l}\leq \mathcal{N}$ and $0\leq m_{i,\triangle}^{s},m_{j,\nabla}^{r},l^{b}_{\triangle},l^{c}_{\nabla} \leq1$.  We have introduced a truncation $\mathcal{N}$ to the occupation number of boson in each lattice, which is practically reasonable since the number of bosons available in an experiment is always limited.
For an illustration, the vacuum of the free theory is given by
\begin{eqnarray}\label{vac}
|Vac\rangle=\otimes_{\vec{p}\in\Gamma} | 0^{a}_{l},0^{s}_{i,\triangle},0^{r}_{j,\nabla},0^{b}_{\triangle},0^{c}_{\nabla} \rangle_{\vec{p}}.
\end{eqnarray}
The states with different number of free particles can be related to each other by the creation and annihilation operators
\begin{eqnarray}  \nonumber
&&\bar{a}^{a\dagger}_{\vec{p},l}|n^{a}_{l}\rangle_{\vec{p}}=\sqrt{n+1}|(n+1)^{a}_{l}\rangle_{\vec{p}},\ \ \bar{a}^{a\dagger}_{\vec{p},l}|\mathcal{N}^{a}_{l}\rangle_{\vec{p}}=0,\\   \nonumber
&&\bar{a}^{a}_{\vec{p},l}|n^{a}_{l}\rangle_{\vec{p}}=\sqrt{n}|(n-1)^{a}_{l}\rangle_{\vec{p}},\ \ \bar{a}^{a}_{\vec{p},l}|0^{a}_{l}\rangle_{\vec{p}}=0,\\  \nonumber
&&b^{s\dagger}_{\vec{p},i}|0^{s}_{i,\triangle}\rangle_{\vec{p}}=|1^{s}_{i,\triangle}\rangle_{\vec{p}}, \ \ b^{s\dagger}_{\vec{p},i}|1^{s}_{i,\triangle}\rangle_{\vec{p}}=0,\\  \nonumber
&&b^{s}_{\vec{p},i}|1^{s}_{i,\triangle}\rangle_{\vec{p}}=|0^{s}_{i,\triangle}\rangle_{\vec{p}}, \ \ b^{s}_{\vec{p},i}|0^{s}_{i,\triangle}\rangle_{\vec{p}}=0,\\   \nonumber
&&c^{s\dagger}_{\vec{p},i}|0^{s}_{i,\nabla}\rangle_{\vec{p}}=|1^{s}_{i,\nabla}\rangle_{\vec{p}}, \ \ c^{s\dagger}_{\vec{p},i}|1^{s}_{i,\nabla}\rangle_{\vec{p}}=0,\\  \nonumber
&&c^{s}_{\vec{p},i}|1^{s}_{i,\nabla}\rangle_{\vec{p}}=|0^{s}_{i,\nabla}\rangle_{\vec{p}}, \ \ b^{s}_{\vec{p},i}|0^{s}_{i,\triangle}\rangle_{\vec{p}}=0,\\   \nonumber
&&d^{a\dagger}_{\vec{p}}|0^{a}_{\triangle}\rangle_{\vec{p}}=|1^{a}_{\triangle}\rangle_{\vec{p}}, d^{a\dagger}_{\vec{p}}|1^{a}_{\triangle}\rangle_{\vec{p}}=0,\\  \nonumber
&&d^{a}_{\vec{p}}|1^{a}_{\triangle}\rangle_{\vec{p}}=|0^{a}_{\triangle}\rangle_{\vec{p}}, d^{a}_{\vec{p}}|0^{a}_{\triangle}\rangle_{\vec{p}}=0,\\  \nonumber
&&e^{a}_{\vec{p}}|1^{a}_{\nabla}\rangle_{\vec{p}}=|0^{a}_{\nabla}\rangle_{\vec{p}}, e^{a}_{\vec{p}}|0^{a}_{\nabla}\rangle_{\vec{p}}=0,\\  
&&e^{a\dagger}_{\vec{p}}|0^{a}_{\nabla}\rangle_{\vec{p}}=|1^{a}_{\nabla}\rangle_{\vec{p}}, e^{a\dagger}_{\vec{p}}|1^{a}_{\nabla}\rangle_{\vec{p}}=0,
\end{eqnarray}
where $\bar{a}^{a\dagger}_{\vec{p},l}$ and $\bar{a}^{a}_{\vec{p},l}$ are the truncated creation and annihilation operators of bosons, 
$b^{s\dagger}_{\vec{p},i}$, $c^{s\dagger}_{\vec{p},i}$,  $b^{s}_{\vec{p},i}$, $c^{s}_{\vec{p},i}$ are creation and annihililation operators for the fermions, and $d^{a\dagger}_{\vec{p}}$, $e^{a\dagger}_{\vec{p}}$, $d^{a}_{\vec{p}}$, $e^{a}_{\vec{p}}$ for ghosts.

According to \cite{RolandoSomma2003Quantum}, the one-to-one map for gauge boson states to qubits can be obtained 
\begin{eqnarray}  \nonumber
|0^{a}_{l}\rangle_{\vec{p}} &\leftrightarrow& |\uparrow_{0} \downarrow_{1}\downarrow_{2}\cdots \downarrow_{\mathcal{N}} \rangle_{\vec{p},a,l},\\  \nonumber
|1^{a}_{l}\rangle_{\vec{p}} &\leftrightarrow& |\downarrow_{0} \uparrow_{1}\downarrow_{2}\cdots \downarrow_{\mathcal{N}} \rangle_{\vec{p},a,l},\\  \nonumber
&\vdots&\\     \label{bosonmap}
|{\mathcal{N}}_{l}^{a}\rangle_{\vec{p}} &\leftrightarrow& |\downarrow_{0} \downarrow_{1}\downarrow_{2}\cdots \uparrow_{\mathcal{N}} \rangle_{\vec{p},a,l}.
\end{eqnarray}
The mapping between fermion states and qubits are
\begin{eqnarray}\label{fermionmap}
|0^{s}_{i,\triangle}\rangle_{\vec{p}} \leftrightarrow |\uparrow\rangle_{\vec{p},i,s,\triangle}, &&
|1^{s}_{i,\triangle}\rangle_{\vec{p}} \leftrightarrow |\downarrow\rangle_{\vec{p},i,s,\triangle},\\
|0^{s}_{i,\nabla}\rangle_{\vec{p}} \leftrightarrow |\uparrow\rangle_{\vec{p},i,s,\nabla}, &&
|1^{s}_{i,\nabla}\rangle_{\vec{p}} \leftrightarrow |\downarrow\rangle_{\vec{p},i,s,\nabla},
\end{eqnarray}
and ghost fields and qubits are related by
\begin{eqnarray}
|0^{a}_{\triangle}\rangle_{\vec{p}} \leftrightarrow |\uparrow\rangle_{\vec{p},a,\triangle}, &&
|1^{a}_{\triangle}\rangle_{\vec{p}} \leftrightarrow |\downarrow\rangle_{\vec{p},a,\triangle},\\
|0^{a}_{\nabla}\rangle_{\vec{p}} \leftrightarrow |\uparrow\rangle_{\vec{p},a,\nabla}, &&
|1^{a}_{\nabla}\rangle_{\vec{p}} \leftrightarrow |\downarrow\rangle_{\vec{p},a,\nabla}.
\end{eqnarray}

Summing up all qubits which expand the Fock space for one-flavor $SU(N)$ Yang-Mills theory, the total number of the working qubits is $[2(N^{2}-1)(\mathcal{N}+1)+4N]\mathcal{V}$. The space complexity of working qubits is polynomial.

We use Jordan-Wigner mapping to give matrix representations to the creation and annihilation operators
\begin{eqnarray} \label{JW} \nonumber
\bar{a}^{a\dagger}_{\vec{p},l}=\sum^{N}_{h=0}\sqrt{h+1}&&\sigma^{h,\vec{p},a,l}_{-}\sigma^{h+1,\vec{p},a,l}_{+},\\ 
\bar{a}^{a}_{\vec{p},l}=\sum^{N}_{h=0}\sqrt{h+1}&&\sigma^{h,\vec{p},a,l}_{+}\sigma^{h+1,\vec{p},a,l}_{-},\nonumber\\ 
b^{s\dagger}_{\vec{p},i}=(\prod_{\alpha_{1}} -\sigma^{\alpha_{1}}_{z})\sigma^{\vec{p},i,s,\triangle}_{+}, &&b^{s}_{\vec{p},i}=(\prod_{\alpha_{1}} -\sigma^{\alpha_{1}}_{z})\sigma^{\vec{p},i,s,\triangle}_{-},\label{b2} \nonumber \\ 
c^{s\dagger}_{\vec{p},i}=(\prod_{\alpha_{2}} -\sigma^{\alpha_{2}}_{z})\sigma^{\vec{p},i,s,\nabla}_{+}, &&c^{s}_{\vec{p},i}=(\prod_{\alpha_{2}} -\sigma^{\alpha_{2}}_{z})\sigma^{\vec{p},i,s,\nabla}_{-},\label{b2} \nonumber \\ 
d^{a\dagger}_{\vec{p}}=(\prod_{\alpha_{3}} -\sigma^{\alpha_{3}}_{z})\sigma^{\vec{p},a,\triangle}_{+}, && d^{a}_{\vec{p}}=(\prod_{\alpha_{3}} -\sigma^{\alpha_{3}}_{z})\sigma^{\vec{p},a,\triangle}_{-}, \nonumber \\
e^{a\dagger}_{\vec{p}}=(\prod_{\alpha_{4}} -\sigma^{\alpha_{4}}_{z})\sigma^{\vec{p},a,\nabla}_{+}, &&e^{a}_{\vec{p}}=(\prod_{\alpha_{4}} -\sigma^{\alpha_{4}}_{z})\sigma^{\vec{p},a,\nabla}_{-},
\end{eqnarray}
with $\sigma_{\pm}=\frac{1}{2}(\sigma_{x}\pm i\sigma_{y})$ and 
\begin{eqnarray}
\alpha_{1} \in \{ qubit | \mathcal{K}(qubit)<\mathcal{K}(\vec{p},i,s,\triangle)\},\\
\alpha_{2} \in \{ qubit| \mathcal{K}(qubit)<\mathcal{K}(\vec{p},i,s,\nabla)\},\\
\alpha_{3} \in \{ qubit| \mathcal{K}(qubit)<\mathcal{K}(\vec{p},a,\triangle)\},\\
\alpha_{4}\in \{ qubit| \mathcal{K}(qubit)<\mathcal{K}(\vec{p},a,\nabla)\},
\end{eqnarray}
where $\mathcal{K}(qubit)$ is an integer and is the primary key of the qubit. The mapping digitally expresses the action of the creation and annihilation operators to the particle states by the operation of the Pauli matrices $\sigma_{x,y,z}$ on the qubits, with every fermionic operator paying the cost for $O(n)$ qubit operations. In the interaction picture, the Hamiltonian is composed of creation and annihilation operators, which makes the simulation of the evolution of the particle states in a straightforward manner. Note that there are other alternate methods of transforming operators to matrices, one of which commonly used is known as the Bravyi-Kitaev transformation \cite{Bravyi2004Universal, Seeley2012}. As compared to the Jordan-Wigner mapping, it can reduce the simulation cost of qubit operations from $O(n)$ to $O(logn)$ for one fermionic operation. 
 
\subsection{Trotter decomposition of Dyson series}
The evolution of the particle state in the interaction picture is
\begin{eqnarray}\label{Psit}
|\Psi(t)\rangle=U(t,t_{0})|\Psi(t_{0})\rangle,
\end{eqnarray}
where $|\Psi(t_{0})\rangle$ is the initial state. The  $U(t,t_{0})$ is the evolution operator (Dyson series)
\begin{eqnarray}
U(t,t)=I, &&\ U(t,t_{0})=U(t,t^{\prime})U(t^{\prime},t_{0}).
\end{eqnarray}

Dividing the time interval into slices of duration $\Delta t=(t-t_{0})/n$, we have
\begin{eqnarray}\label{U}   \nonumber
U(t,t_{0})&=&U(t,t_0+(n-1)\Delta t) \cdots U(t_{0}+\Delta t,t_{0}).
\end{eqnarray}
When $\Delta t$ is small enough, we have
\begin{eqnarray}\label{UDelta}
U(t+\Delta t,t )=e^{-iH_{I}(t)  \Delta t}
\end{eqnarray}
with $H_I$ the interaction part of the Hamiltonian
\begin{eqnarray}
H_{I}=\int d^{3}x \mathcal{H}_{I}.
\end{eqnarray}
The Hamiltonian density of Yang-Mills theory $\mathcal{H}_{I}$ is a sum of four types of Hermitian interactions: fermion-boson $\mathcal{H}_{FI}$, four-boson $\mathcal{H}_{G4I}$, three-boson $\mathcal{H}_{G3I}$ and ghost-boson $\mathcal{H}_{FPI}$
\begin{eqnarray} \nonumber
\mathcal{H}_{I}&=&\mathcal{H}_{FI}+\mathcal{H}_{G4I}+\mathcal{H}_{G3I}+\mathcal{H}_{FPI}\\   \nonumber
&=&-gA^{a}_{\mu}\bar{\psi}_{i}\gamma^{\mu}t^{a}_{ij}\psi_{j}+\frac{1}{4}g^{2}(f^{eab}A^{a}_{\mu}A^{b}_{\nu})(f^{ecd}A^{\mu c}A^{\nu d})\\  \nonumber
&&+g f^{abc}(\partial_{\mu}A^{a}_{\nu})A^{\mu b}A^{\nu c}
-g f^{abc}(\partial^{\mu} \bar{c}^{a})A^{b}_{\mu}c^{c}.
\end{eqnarray}

Using the Trotter formula \cite{Chuang2015Quantum}  
\begin{eqnarray}
e^{i(A+B)\Delta t}=e^{iA \Delta t} e^{i B \Delta t}+O(\Delta t^{2})
\end{eqnarray}
where $A$ and $B$ are Hermitian operators and taking $\Delta t$ as a small enough quantity so that the terms proportional to $\Delta t^2$ can be ignored, we have
\begin{eqnarray*}
e^{-iH_{I}(t)  \Delta t}=e^{-iH_{FI}  \Delta t} e^{-iH_{G4I}  \Delta t}e^{-iH_{G3I} \Delta t} e^{-iH_{FPI}  \Delta t}.
\end{eqnarray*}
The task following is to substitute (\ref{qfields}) and (\ref{JW}) into the above equation. With repeated use of the Trotter formula, it can be written as continued product of the exponential of the Pauli matrices. The operation of each factor of the product on qubits can be implemented by a quantum circuit. For illustration, we take $e^{-iH_{FI}\Delta t}$ as an example to show the procedure. 

With (\ref{qfields}), the Hamiltonian $H_{FI}$ can be written as
\begin{eqnarray*} \nonumber
H_{FI}=-g\int d^{3}x (A^{a+}_{\mu}\bar{\psi}^{+}_{i}\gamma^{\mu}t^{a}_{ij}\psi^{+}_{j}+A^{a+}_{\mu}\bar{\psi}^{+}_{i}\gamma^{\mu}t^{a}_{ij}\psi^{-}_{j}\\
+A^{a+}_{\mu}\bar{\psi}^{-}_{i}\gamma^{\mu}t^{a}_{ij}\psi^{+}_{j}+A^{a+}_{\mu}\bar{\psi}^{-}_{i}\gamma^{\mu}t^{a}_{ij}\psi^{-}_{j}+H.c.),
\end{eqnarray*}
where the superscripts $+$ and $-$ on the fields denote the positive and negative frequency components. We pick out a typical structure in $H_{FI}$ 
\begin{eqnarray} \nonumber
H_{1}&=&-g\int d^{3}x (A^{a+}_{\mu}\bar{\psi}^{-}_{i}\gamma^{\mu}t^{a}_{ij}\psi^{+}_{j}+H.c.).
\end{eqnarray}
Discretizing the momentum space and applying the gauge boson occupation number truncation, in the case $\mathcal{K}(\vec{p}_{2},i,r,\triangle) \leq  \mathcal{K}(\vec{p}_{3},j,s,\triangle)$, applying the Jordan-Wigner mappings (\ref{JW}) to the terms, we have the Hamiltonian in terms of the Pauli matrices
\begin{widetext}
\begin{eqnarray}   \label{H3}  \nonumber
H_{1}&=&-g \sum_{\vec{p}_{1},\vec{p}_{2},\vec{p}_{3}\in \Gamma}
\frac{\kappa^{3}_{1}\kappa^{3}_{2}\kappa^{3}_{3}}{(2\pi)^{6}\sqrt{8\omega_{\vec{p}_{1}}\omega_{\vec{p}_{2}}\omega_{\vec{p}_{3}}}}\delta^{(3)}(\vec{p}_{1}-\vec{p}_{2}+\vec{p}_{3})  \sum_{l,r,s}[\bar{a}^{a}_{\vec{p}_{1},l}b^{r\dagger}_{\vec{p}_{2},i}b^{s}_{\vec{p}_{3},j}\bar{u}^{r}_{p_{2}} (\epsilon_{l}\cdot\gamma) t^{a}_{ij}u^{s}_{p_{3}}e^{-i\omega_{1} t}\!+\!H.c.]\\ \nonumber
&=&-g \sum_{\vec{p}_{1},\vec{p}_{2},\vec{p}_{3}\in \Gamma}
\frac{\kappa^{3}_{1}\kappa^{3}_{2}\kappa^{3}_{3}}{(2\pi)^{6}\sqrt{8\omega_{\vec{p}_{1}}\omega_{\vec{p}_{2}}\omega_{\vec{p}_{3}}}}\delta^{(3)}(\vec{p}_{1}-\vec{p}_{2}+\vec{p}_{3}) \frac{1}{8}\sum_{l,r,s}
\sum_{h=0}^{\mathcal{N}}  \left[\left(\sigma^{h,\vec{p}_{1},a,l}_{x}\sigma^{h+1,\vec{p}_{1},a,l}_{x}+ \sigma^{h,\vec{p}_{1},a,l}_{y}\sigma^{h+1,\vec{p}_{1},a,l}_{y} \right)\right.  \\        \nonumber
&&\left(   (-\sigma^{\vec{p}_{2},i,r,\triangle}_{y})(\prod_{\alpha}-\sigma^{\alpha}_{z} ) \sigma^{\vec{p}_{3},j,s,\triangle}_{x}W^{a,r,s}_{2,ij,l}+\sigma^{\vec{p}_{2},i,r,\triangle}_{y}(\prod_{\alpha}-\sigma^{\alpha}_{z} ) \sigma^{\vec{p}_{3},j,s,\triangle}_{y}W^{a,r,s}_{1,ij,l} \right.\\    \nonumber
&&\left.+\sigma^{\vec{p}_{2},i,r,\triangle}_{x}(\prod_{\alpha}-\sigma^{\alpha}_{z} ) \sigma^{\vec{p}_{3},j,s,\triangle}_{x}W^{a,r,s}_{1,ij,l} 
+\sigma^{\vec{p}_{2},i,r,\triangle}_{x}(\prod_{\alpha}-\sigma^{\alpha}_{z} ) \sigma^{\vec{p}_{3},j,s,\triangle}_{y}W^{a,r,s}_{2,ij,l}    \right)\\  \nonumber
&&+\left(\sigma^{h,\vec{p}_{1},a,l}_{y}\sigma^{h+1,\vec{p}_{1},a,l}_{x} -\sigma^{h,\vec{p}_{1},a,l}_{x}\sigma^{h+1,\vec{p}_{1},a,l}_{y}\right)   \\   \nonumber
&&\left( (-\sigma^{\vec{p}_{2},i,r,\triangle}_{y})(\prod_{\alpha}-\sigma^{\alpha}_{z} ) \sigma^{\vec{p}_{3},j,s,\triangle}_{x}W^{a,r,s}_{1,ij,l}-\sigma^{\vec{p}_{2},i,r,\triangle}_{y}(\prod_{\alpha}-\sigma^{\alpha}_{z} ) \sigma^{\vec{p}_{3},j,s,\triangle}_{y}W^{a,r,s}_{2,ij,l} \right.  \\   \nonumber
&&\left. \left.-\sigma^{\vec{p}_{2},i,r,\triangle}_{x}(\prod_{\alpha}-\sigma^{\alpha}_{z} ) \sigma^{\vec{p}_{3},j,s,\triangle}_{x}W^{a,r,s}_{2,ij,l} 
+\sigma^{\vec{p}_{2},i,r,\triangle}_{x}(\prod_{\alpha}-\sigma^{\alpha}_{z} ) \sigma^{\vec{p}_{3},j,s,\triangle}_{y}W^{a,r,s}_{1,ij,l}    \right)\right],
\end{eqnarray}
\end{widetext}
{where $\omega_{1}=\omega_{\vec{p}_1}-\omega_{\vec{p}_2}+\omega_{\vec{p}_3}$ and $\delta^{(3)}$  being the three-dimensional discrete delta function. The $\kappa_i$ denotes the spacing of the momentum $\vec{p}_i$ lattice, the repeated indices in the square bracket is understood as the summation convention and $\alpha \in \{ (\vec{p}^{\prime}, i^{\prime}, s^{\prime},\triangle)|   \mathcal{K}(\vec{p}_{2},i,r,\triangle) < \mathcal{K}(\vec{p}^{\prime}, i^{\prime}, s^{\prime},\triangle) <  \mathcal{K}(\vec{p}_{3},j,s,\triangle) \}$.  The $W^{a,r,s}_{1,ij,l},W^{a,r,s}_{2,ij,l} \in \mathbb{R}$ are defined by
\begin{eqnarray}
\bar{u}^{r}(p_{2}) (\epsilon_{l}\cdot\gamma) t^{a}_{ij}u^{s}(p_{3})e^{-i\omega_{1} t}=W^{a,r,s}_{1,ij,l}+i W^{a,r,s}_{2,ij,l}.
\end{eqnarray}
It is showed that the $H_{1}$ is sum of terms which are products of Pauli matrices
\begin{eqnarray}
H_{\xi}=\lambda_{\xi}\otimes_{r=1}^{M} \sigma^{r}_{c(r)}, H_{1}=\sum_{\xi}H_{\xi},
\end{eqnarray}
where $\lambda_{\xi}\in \mathbb{R}$ is the effective coupling strength of $H_{\xi}$. The Pauli matrix $\sigma^{r}_{c(r)}$ acts on the $r$th qubit, with $c(r)\in \{0,1,2,3\}$ and $\sigma_{c(r)}= (I,\sigma_{x},\sigma_{y},\sigma_{z} )$. 
With repeated use of the Trotter formula, it is easy to have
\begin{eqnarray}
e^{-iH_{1}\Delta t}=\prod_{\xi} e^{-i H_{\xi} \Delta t},
\end{eqnarray}
and the similar expression for $e^{-iH_{FI}\Delta t}$ as well.  The $H_\xi$ is Hermitian and thus $e^{-iH_{\xi}\Delta t}$ can be simulated by quantum circuits. Summing up all the indices, $\vec{p}_{1}$, $\vec{p}_{2}$, $\vec{p}_{3}$, $a$, $r$, $s$, $i$, $j$, $l$, related CNOT operations in $H_{1}$ to simulate $e^{-iH_{1}\Delta t}$, the CNOT operations is less than  $512(2+\mathcal{V}N)(N^{2}-1)N^{2}(\mathcal{N}+1)$. 
Similarly, to simulate $e^{-iH_{FI}\Delta t}$, we need less than  
$64\mathcal{V}\left[(1+N)^{2}(\mathcal{V}^{2}-3)+64\mathcal{V}\right]N^{2}(N^{2}-1)(\mathcal{N}+1) $ CNOT operations. 
}

As an unitary transformation, the action of the matrix $e^{-iH_{1}\Delta t}$ to the qubits can be effectively implemented by a quantum circuits \cite{Chuang2015Quantum}. This is an advantage of the quantum simulation over the classical one which performs the calculation by perturbation in expanding the exponential into power series when the coupling is in the weak limit or by the lattice gauge theory to numerically evaluate the functional integral in the strong coupling limit.

\begin{figure}
\begin{center}
\includegraphics[width=80 mm]{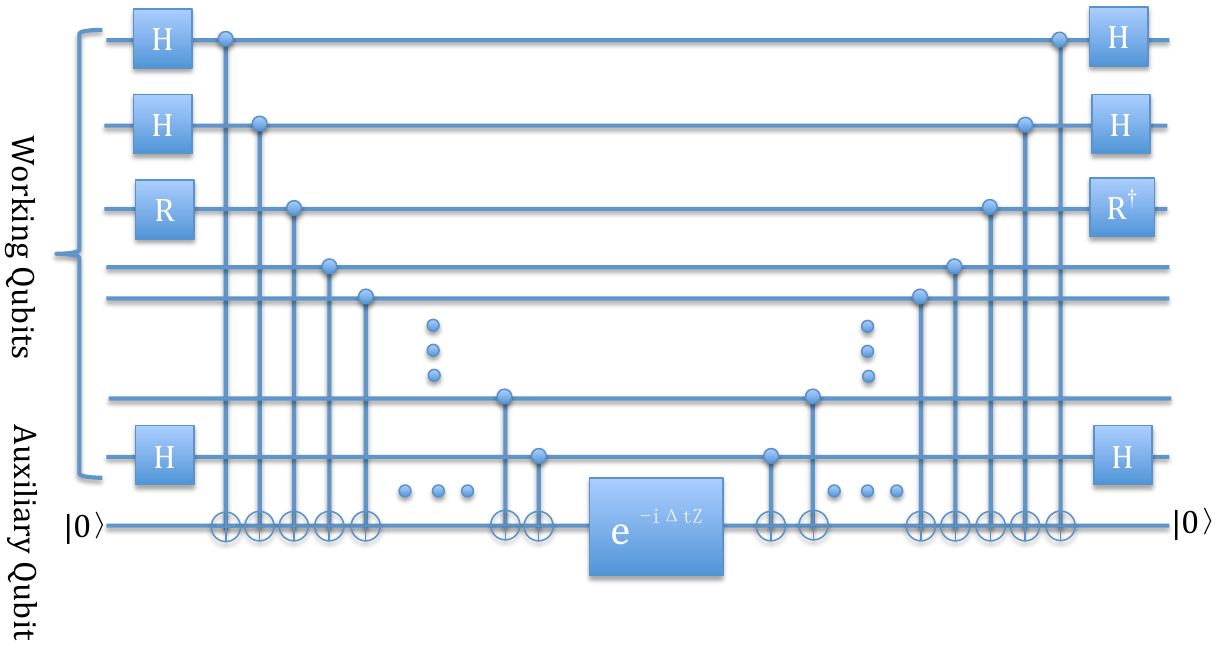}
\caption{\label{fig1}Quantum circuits which perform unitary transformation $e^{-i\Delta t \sigma^{1}_{x}\sigma^{2}_{x}\sigma^{3}_{y}(\prod_{\alpha}\sigma^{\alpha}_{z} ) \sigma^{r}_{x} }$ on working qubits.  The symbol $H$ means Hadamard operation. The $Z$ is the $\sigma_z$ matrix and the $R$ is a transformation: $R\sigma_{y}R^{\dagger}=\sigma_z$. This unitary transformation is one part of transformation in $e^{-i H_{\xi} \Delta t}$. There are lots of CNOT operations and single qubit operations being used in this quantum circuits. But only one auxiliary qubit is needed.}
\end{center}
\end{figure}

An example of quantum circuit to simulate the $e^{-i\Delta t \sigma^{1}_{x}\sigma^{2}_{x}\sigma^{3}_{y}(\prod_{\alpha}\sigma^{\alpha}_{z} ) \sigma^{r}_{x} }$ operation on working qubits is given in Fig.~\ref{fig1} where the symbol $H$ denotes Hadamard operation and $R$  can be chosen as
\begin{eqnarray}
R=\frac{1}{\sqrt{2}}\left(\begin{array}{cc}1 & -i \\ -i & 1\end{array}\right),
\end{eqnarray}
and we have $R\sigma_{y}R^{\dagger}=\sigma_z$.  The unitary transformation is one part of transformation in $e^{-i H_{\xi} \Delta t}$. 
With the same treament, the operations of $e^{-i H_{G4I}\Delta t}$, $e^{-i H_{G3I}\Delta t}$ and $e^{-i H_{FBI}\Delta t}$ can be implemented by quantum circuits, which add up to the quantum simulation of the evolution (\ref{Psit}).  These quantum circuits are constructed by a vast quantity of CNOT operations, single qubit operations and one auxiliary qubit. Summing up the CNOT operations used in quantum circuits, the time complexity to simulate $e^{-iH_{FI}\Delta t}$, $e^{-i H_{G4I}\Delta t}$, $e^{-i H_{G3I}\Delta t}$ and $e^{-iH_{FPI}  \Delta t}$ is lower than $O(N^{6}\mathcal{V}^{3}(\mathcal{N}+1))$, $O(N^{10}\mathcal{V}^{3}(\mathcal{N}+1)^{4})$, $O(N^{6}\mathcal{V}^{2}(\mathcal{N}+1)^{3})$ and $O(N^{6}\mathcal{V}^{3}(\mathcal{N}+1))$ respectively. This leads to the time complexity of the simulation $U(t,t_{0})$ to be $O(N^{10}\mathcal{V}^{3}(\mathcal{N}+1)^{4}n)$.

\section{Potential Applications}
\label{application}

A good venue to make use of the algorithm is QCD where simulation can be performed to study the hadron physics in the future when the quantum computer becomes full-developed. One interesting application might be the visualization of the hadronization, by giving dynamic information on how the unobservable free quarks form into a hadron. For demonstration purpose, we sketch out examples of meson and baryon with one flavor.
In interaction picture, the initial states are free particles. The free quarks are unavailable in reality, but they can be prepared in quantum computers. 

The first example is the formation of meson. The initial state to evolve is the three-colour superposition of a pair of free quark and anti-quark with equal amplitude
\begin{eqnarray}
|\Psi(t_{0})\rangle =\frac{1}{\sqrt{3}}\sum^{3}_{i=1} |q^{s_1}_{\vec{p}_{1},i} \bar{q}^{s_2}_{\vec{p}_{2},i}\rangle,
\end{eqnarray}
where $\bar{q}^{s}_{\vec{p},i}$ ($q^{s}_{\vec{p},i}$) denotes anti-quark (quark) with momentum $\vec{p}$, spin $s$ and color $i$. 
To prepare the color singlet initial state, one can first produce, according to Eq.~(\ref{fermionmap}), two qubits to represent the two-quark state 
\begin{eqnarray}
|q^{s_1}_{\vec{p}_{1},1} \bar{q}^{s_2}_{\vec{p}_{2},1}\rangle. 
\end{eqnarray}
Then with a specific unitary transformation $\cal T$ acting on the qubits, one can have the qubit representation of the initial state
\begin{eqnarray}
|\Psi(t_{0})\rangle={\cal T} |q^{s_1}_{\vec{p}_{1},1} \bar{q}^{s_2}_{\vec{p}_{2},1}\rangle. 
\end{eqnarray}
The operation of the transformation $\mathcal{T}$ can be implemented by quantum circuits and one specific $\mathcal{T}$ is given in \ref{A2}. The complexity to prepare initial state is $O(1)$.
At last, one can turn on the quantum circuits and adjust the value of the coupling constant $g$ in the evolution operator $U(t,t_0)$ to accomplish the simulation of the evolution from the initial state to a meson
\begin{eqnarray}
U(t,t_{0})|\Psi(t_{0})\rangle  \rightarrow  |Meson\rangle.
\end{eqnarray}

The simulation of baryon can be perform in a similar way. The initial state are three quarks with different color
\begin{eqnarray}
|\tilde{\Psi}(t_{0})\rangle = |q^{s_1}_{\vec{p}_{1},1} q^{s_2}_{\vec{p}_{2},2}q^{s_3}_{\vec{p}_{3},3}\rangle.
\end{eqnarray}
Acting $U(t,t_0)$ to the initial state, the hadronization of the baryon can be simulated
\begin{eqnarray}
U(t,t_{0})|\tilde{\Psi}(t_{0})\rangle  \rightarrow  |Baryon\rangle.
\end{eqnarray}

In the simulation, one can extract information on the behavior of the particles at any k-slice discrete time $t=t_{0}+k \Delta t$ from the evolving wave function $|\Psi(t)\rangle$, with $k=0,1\cdots n$. A typical and straightforward measurement is the possibility $P(n_{q},\vec{p},t)$ of each kind of particles with occupation number $n_{q}$ at momentum $\vec{p}$, by counting the number of the corresponding qubit collapsing into spin $|\uparrow\rangle$ or $|\downarrow\rangle$ state.

\section{Discussion and Conclusion}
\label{con}
This digital algorithm aims at using fully developed quantum computer to simulate Yang-Mills theory in interaction picture. We use the canonical quantization formalism and perform the discretization in momentum space in order to avoid the fermion doubling problem. The spins of fermions and gauge bosons are seriously treated. The momentum lattice Yang-Mills theory is gauge invariant and can be made ghostfree when adopting the axial or the light-cone gauge. 

Though this algorithm may help us understand the low energy QCD in the future when quantum computer is powerful enough, it is not practical at moment for the unreachable required amount of quantum devices. Taking hadronization simulation as an example, the minimal size of the algorithm with physical meaning might be $N=3$, $\mathcal{V}=3^{3}$ and $\mathcal{N}=1$. The requisite qubit number is $[2(N^{2}-1)(\mathcal{N}+1)+4N]\mathcal{V}+1=1189$. The CNOT operations is about $655360 n_{t}(N^{2}-1)^{5}\mathcal{V}^{3}(\mathcal{N}+1)^{4}\approx 6.7n_{t}\times 10^{15}$, with $n_{t}$ the number of $\Delta t$ steps. Such a large amount of CNOT operations imposes a stringent requirement on the fidelity. If the total error is supposed to be less than $10\%$ without using error correction qubits in the simulation, the fidelity of each CNOT operation should require to be higher than $0.9^{\frac{1}{6.7n_{t}\times 10^{15}}}$. 

In summary, we have presented a quantum algorithm to non-perturbatively simulate the Yang-Mills theory. Both the space complexity and time complexity of this algorithm is polynomial. This efficient algorithm may pave the way to study the physics in and beyond the Standard Model of particle physics with quantum computers in the future. 

\begin{acknowledgements}
This work is supported by National Basic Research Program of China under Grant No. 2016YFA0301903, and the National Natural Science Foundation of China under Grants Nos. 11174370, 11304387, 61632021, 11305262, 61205108, 11475258 and 11574398.
\end{acknowledgements}

\bibliographystyle{apsrev4-1}


%

\newpage
\section*{Appendix A: The Expression for Four-Gluon Interaction}
The decomposition of four gluon interaction is rather complicated. We use a term $H_2$ in $H_{G4I}$ as an example 
\begin{eqnarray}
H_{2}=\frac{g^{2}}{4}\int d^{3}x f^{eab}f^{ecd}A^{a+}_{\mu}A^{b+}_{\nu}A^{\mu c-}A^{\nu d-}.
\end{eqnarray}
After performing the discretization and gluon occupation number truncation, one has
\begin{eqnarray}  \label{H2}  \nonumber
H_{2}&=&\frac{g^{2}}{4} \sum_{\vec{p}_{1}\in \Gamma}\sum_{\vec{p}_{2}\in \Gamma}\sum_{\vec{p}_{3}\in \Gamma}\sum_{\vec{p}_{4}\in \Gamma}
\frac{\kappa^{3}_{1}\kappa^{3}_{2}\kappa^{3}_{3}a^{3}_{4}}{4(2\pi)^{9}\sqrt{\omega_{\vec{p}_{1}}\omega_{\vec{p}_{2}}\omega_{\vec{p}_{3}}\omega_{\vec{p}_{4}}}}\\  
&&\delta^{(3)}(\vec{p}_{1}+\vec{p}_{2}-\vec{p}_{3}-\vec{p}_{4})f^{eab}f^{ecd}\\  \nonumber
&&\sum_{l_{1},l_{2},l_{3},l_{4}} \bar{a}^{a}_{\vec{p}_{1},l_{1}}\bar{a}^{b}_{\vec{p}_{2},l_{2}}\bar{a}^{c\dagger}_{\vec{p}_{3},l_{3}}\bar{a}^{d\dagger}_{\vec{p}_{4},l_{4}} \epsilon_{\mu,l_{1}}\epsilon_{\nu,l_{2}}\epsilon^{\mu*}_{l_{3}}\epsilon^{\nu*}_{l_{4}} e^{-i \omega_{2}t},
\end{eqnarray}
with $\omega_{2}=\omega_{\vec{p}_{1}}+\omega_{\vec{p}_{2}}-\omega_{\vec{p}_{3}}-\omega_{\vec{p}_{4}}$. We calculate the factor $I_2$ in $H_{2}$ 
\begin{eqnarray}  \nonumber \label{I2}  
I_{2}&=&
\frac{1}{2}\sum_{l_{1},l_{2},l_{3},l_{4}}
\left(\bar{a}^{a}_{\vec{p}_{1},l_{1}}\bar{a}^{b}_{\vec{p}_{2},l_{2}}\bar{a}^{c\dagger}_{\vec{p}_{3},l_{3}}\bar{a}^{d\dagger}_{\vec{p}_{4},l_{4}} \epsilon_{\mu,l_{1}}\epsilon_{\nu,l_{2}}\epsilon^{\mu*}_{l_{3}}\epsilon^{\nu*}_{l_{4}} e^{-i \omega_{2}t}\right.\\
&&+\left.\bar{a}^{d}_{\vec{p}_{4},l_{4}}\bar{a}^{c}_{\vec{p}_{3},l_{3}} \bar{a}^{b\dagger}_{\vec{p}_{2},l_{2}} \bar{a}^{a\dagger}_{\vec{p}_{1},l_{1}} \epsilon^{*}_{\mu,l_{1}}\epsilon^{*}_{\nu,l_{2}}\epsilon^{\mu}_{l_{3}}\epsilon^{\nu}_{l_{4}} e^{i \omega_{2}t}\right),
\end{eqnarray}
and give its expression in two cases categorized by the quantum numbers of the particles. 

For the case 
\begin{eqnarray}\label{p4} 
&&(\vec{p}_{1},a,l_{1})=(\vec{p}_{3},c,l_{3})\neq (\vec{p}_{2},b,l_{2})=(\vec{p}_{4},d,l_{4})\nonumber\\   \label{p5}  
or\ &&(\vec{p}_{1},a,l_{1})=(\vec{p}_{4},d,l_{4})\neq(\vec{p}_{2},b,l_{2})=(\vec{p}_{3},c,l_{3}),
\end{eqnarray}
we have 
\begin{eqnarray}   \label{I21}
I_{2}&=&\sum_{l_{1},l_{2},l_{3},l_{4}}\bar{a}^{a}_{\vec{p}_{1},l_{1}}\bar{a}^{c\dagger}_{\vec{p}_{3},l_{3}}\bar{a}^{b}_{\vec{p}_{2},l_{2}}\bar{a}^{d\dagger}_{\vec{p}_{4},l_{4}}W_{3,l_{1},l_{2},l_{3},l_{4}} ,
\end{eqnarray}
with
\begin{eqnarray*}
&&\bar{a}^{a}_{\vec{p}_{1},l_{1}}\bar{a}^{c\dagger}_{\vec{p}_{3},l_{3}}\\
&=&
\sum^{\mathcal{N}}_{h_{1},h_{2}=0}\sqrt{(h_{1}+1)(h_{2}+1)}\\
&&\frac{1}{16}\left[(\sigma^{h_{1},\vec{p}_{1},a,l_{1}}_{x}\sigma^{h_{1}+1,\vec{p}_{1},a,l_{1}}_{x}
+\sigma^{h_{1},\vec{p}_{1},a,l_{1}}_{y}\sigma^{h_{1}+1,\vec{p}_{1},a,l_{1}}_{y})\right.\\
&&(\sigma^{h_{2},\vec{p}_{3},c,l_{3}}_{x}\sigma^{h_{2}+1,\vec{p}_{3},c,l_{3}}_{x}
+\sigma^{h_{2},\vec{p}_{3},c,l_{3}}_{y}\sigma^{h_{2}+1,\vec{p}_{3},c,l_{3}}_{y})\\
&&+(\sigma^{h_{1},\vec{p}_{1},a,l_{1}}_{x}\sigma^{h_{1}+1,\vec{p}_{1},a,l_{1}}_{y}
-\sigma^{h_{1},\vec{p}_{1},a,l_{1}}_{y}\sigma^{h_{1}+1,\vec{p}_{1},a,l_{1}}_{x})\\
&&\left.\left.(\sigma^{h_{2},\vec{p}_{3},c,l_{3}}_{x}\sigma^{h_{2}+1,\vec{p}_{3},c,l_{3}}_{y}
-\sigma^{h_{2},\vec{p}_{3},c,l_{3}}_{y}\sigma^{h_{2}+1,\vec{p}_{3},c,l_{3}}_{x})\right]\right|_{h_{1}\neq h_{2}}\\
&&+\sum_{h_{1}}^{\mathcal{N}}\frac{h_{1}+1}{4}(I^{h_{1},\vec{p}_{1},a,l_{1}}+\sigma^{h_{1},\vec{p}_{1},a,l_{1}}_{z})(I^{h_{1}+1,\vec{p}_{1},a,l_{1}}-\sigma^{h_{1}+1,\vec{p}_{1},a,l_{1}}_{z}),
\end{eqnarray*}
in which the repeated indices are not summed. 

When the quantum numbers of the bosons $(\vec{p}_{1},a,l_{1})$, $(\vec{p}_{2},b,l_{3})$, $(\vec{p}_{3},c,l_{3})$ and $(\vec{p}_{4},d,l_{4})$ do not equal to each other,  we have another expression
for $I_{2}$
\begin{eqnarray} \label{I22}  \nonumber
I_{2}&=&\frac{1}{256}\sum_{l_{1},l_{2},l_{3},l_{4}}\sum_{h_{1},h_{2},h_{3},h_{4}=0}^{\mathcal{N}}\sqrt{(h_{1}+1)(h_{2}+1)(h_{3}+1)(h_{4}+1)}\\  \nonumber
&&\left[(\Omega_{1}\Omega_{3}-\Omega_{2}\Omega_{4})W_{3,l_{1},l_{2},l_{3},l_{4}}-(\Omega_{2}\Omega_{3}+\Omega_{1}\Omega_{4})W_{4,l_{1},l_{2},l_{3},l_{4}}\right],
\end{eqnarray}
where the $\Omega_{1}$, $\Omega_{2}$, $\Omega_{3}$, $\Omega_{4}$ are 
\begin{eqnarray*}
\Omega_{1}&=&(\sigma^{h_{1},\vec{p}_{1},a,l_{1}}_{x}\sigma^{h_{1}+1,\vec{p}_{1},a,l_{1}}_{x}\sigma^{h_{2},\vec{p}_{2},b,l_{2}}_{x}\sigma^{h_{2}+1,\vec{p}_{2},b,l_{2}}_{x}\\
&&+\sigma^{h_{1},\vec{p}_{1},a,l_{1}}_{x}\sigma^{h_{1}+1,\vec{p}_{1},a,l_{1}}_{x}\sigma^{h_{2},\vec{p}_{2},b,l_{2}}_{y}\sigma^{h_{2}+1,\vec{p}_{2},b,l_{2}}_{y} \\
&&+\sigma^{h_{1},\vec{p}_{1},a,l_{1}}_{y}\sigma^{h_{1}+1,\vec{p}_{1},a,l_{1}}_{y}\sigma^{h_{2},\vec{p}_{2},b,l_{2}}_{x}\sigma^{h_{2}+1,\vec{p}_{2},b,l_{2}}_{x}\\
&&+\sigma^{h_{1},\vec{p}_{1},a,l_{1}}_{y}\sigma^{h_{1}+1,\vec{p}_{1},a,l_{1}}_{y}\sigma^{h_{2},\vec{p}_{2},b,l_{2}}_{y}\sigma^{h_{2}+1,\vec{p}_{2},b,l_{2}}_{y}\\
&&-\sigma^{h_{1},\vec{p}_{1},a,l_{1}}_{x}\sigma^{h_{1}+1,\vec{p}_{1},a,l_{1}}_{y}\sigma^{h_{2},\vec{p}_{2},b,l_{2}}_{x}\sigma^{h_{2}+1,\vec{p}_{2},b,l_{2}}_{y}\\
&&+\sigma^{h_{1},\vec{p}_{1},a,l_{1}}_{x}\sigma^{h_{1}+1,\vec{p}_{1},a,l_{1}}_{y}\sigma^{h_{2},\vec{p}_{2},b,l_{2}}_{y}\sigma^{h_{2}+1,\vec{p}_{2},b,l_{2}}_{x}\\
&&+\sigma^{h_{1},\vec{p}_{1},a,l_{1}}_{y}\sigma^{h_{1}+1,\vec{p}_{1},a,l_{1}}_{x}\sigma^{h_{2},\vec{p}_{2},b,l_{2}}_{x}\sigma^{h_{2}+1,\vec{p}_{2},b,l_{2}}_{y}\\
&&-\sigma^{h_{1},\vec{p}_{1},a,l_{1}}_{y}\sigma^{h_{1}+1,\vec{p}_{1},a,l_{1}}_{x}\sigma^{h_{2},\vec{p}_{2},b,l_{2}}_{y}\sigma^{h_{2}+1,\vec{p}_{2},b,l_{2}}_{x}),
\end{eqnarray*}

\begin{eqnarray*}
\Omega_{2}&=&(-\sigma^{h_{1},\vec{p}_{1},a,l_{1}}_{x}\sigma^{h_{1}+1,\vec{p}_{1},a,l_{1}}_{y}\sigma^{h_{2},\vec{p}_{2},b,l_{2}}_{x}\sigma^{h_{2}+1,\vec{p}_{2},b,l_{2}}_{x}\\
&&-\sigma^{h_{1},\vec{p}_{1},a,l_{1}}_{x}\sigma^{h_{1}+1,\vec{p}_{1},a,l_{1}}_{y}\sigma^{h_{2},\vec{p}_{2},b,l_{2}}_{y}\sigma^{h_{2}+1,\vec{p}_{2},b,l_{2}}_{y} \\
&&+\sigma^{h_{1},\vec{p}_{1},a,l_{1}}_{y}\sigma^{h_{1}+1,\vec{p}_{1},a,l_{1}}_{x}\sigma^{h_{2},\vec{p}_{2},b,l_{2}}_{x}\sigma^{h_{2}+1,\vec{p}_{2},b,l_{2}}_{x}\\
&&+\sigma^{h_{1},\vec{p}_{1},a,l_{1}}_{y}\sigma^{h_{1}+1,\vec{p}_{1},a,l_{1}}_{x}\sigma^{h_{2},\vec{p}_{2},b,l_{2}}_{y}\sigma^{h_{2}+1,\vec{p}_{2},b,l_{2}}_{y}\\
&&-\sigma^{h_{1},\vec{p}_{1},a,l_{1}}_{x}\sigma^{h_{1}+1,\vec{p}_{1},a,l_{1}}_{x}\sigma^{h_{2},\vec{p}_{2},b,l_{2}}_{x}\sigma^{h_{2}+1,\vec{p}_{2},b,l_{2}}_{y}\\
&&-\sigma^{h_{1},\vec{p}_{1},a,l_{1}}_{y}\sigma^{h_{1}+1,\vec{p}_{1},a,l_{1}}_{y}\sigma^{h_{2},\vec{p}_{2},b,l_{2}}_{x}\sigma^{h_{2}+1,\vec{p}_{2},b,l_{2}}_{y}\\
&&+\sigma^{h_{1},\vec{p}_{1},a,l_{1}}_{x}\sigma^{h_{1}+1,\vec{p}_{1},a,l_{1}}_{x}\sigma^{h_{2},\vec{p}_{2},b,l_{2}}_{y}\sigma^{h_{2}+1,\vec{p}_{2},b,l_{2}}_{x}\\
&&+\sigma^{h_{1},\vec{p}_{1},a,l_{1}}_{y}\sigma^{h_{1}+1,\vec{p}_{1},a,l_{1}}_{y}\sigma^{h_{2},\vec{p}_{2},b,l_{2}}_{y}\sigma^{h_{2}+1,\vec{p}_{2},b,l_{2}}_{x}),\\
\end{eqnarray*}

\begin{eqnarray*}
\Omega_{3}&=&(\sigma^{h_{3},\vec{p}_{3},c,l_{3}}_{x}\sigma^{h_{3}+1,\vec{p}_{3},c,l_{3}}_{x}\sigma^{h_{4},\vec{p}_{4},d,l_{4}}_{x}\sigma^{h_{4}+1,\vec{p}_{4},d,l_{4}}_{x}\\
&&+\sigma^{h_{3},\vec{p}_{3},c,l_{3}}_{x}\sigma^{h_{3}+1,\vec{p}_{3},c,l_{3}}_{x}\sigma^{h_{4},\vec{p}_{4},d,l_{4}}_{y}\sigma^{h_{4}+1,\vec{p}_{4},d,l_{4}}_{y}  \\
&& +\sigma^{h_{3},\vec{p}_{3},c,l_{3}}_{y}\sigma^{h_{3}+1,\vec{p}_{3},c,l_{3}}_{y}\sigma^{h_{4},\vec{p}_{4},d,l_{4}}_{x}\sigma^{h_{4}+1,\vec{p}_{4},d,l_{4}}_{x}\\
&&+\sigma^{h_{3},\vec{p}_{3},c,l_{3}}_{y}\sigma^{h_{3}+1,\vec{p}_{3},c,l_{3}}_{y}\sigma^{h_{4},\vec{p}_{4},d,l_{4}}_{y}\sigma^{h_{4}+1,\vec{p}_{4},d,l_{4}}_{y}\\
&&-\sigma^{h_{3},\vec{p}_{3},c,l_{3}}_{y}\sigma^{h_{3}+1,\vec{p}_{3},c,l_{3}}_{x}\sigma^{h_{4},\vec{p}_{4},d,l_{4}}_{y}\sigma^{h_{4}+1,\vec{p}_{4},d,l_{4}}_{x}\\
&&+\sigma^{h_{3},\vec{p}_{3},c,l_{3}}_{y}\sigma^{h_{3}+1,\vec{p}_{3},c,l_{3}}_{x}\sigma^{h_{4},\vec{p}_{4},d,l_{4}}_{x}\sigma^{h_{4}+1,\vec{p}_{4},d,l_{4}}_{y}\\
&&+\sigma^{h_{3},\vec{p}_{3},c,l_{3}}_{x}\sigma^{h_{3}+1,\vec{p}_{3},c,l_{3}}_{y}\sigma^{h_{4},\vec{p}_{4},d,l_{4}}_{y}\sigma^{h_{4}+1,\vec{p}_{4},d,l_{4}}_{x}\\
&&-\sigma^{h_{3},\vec{p}_{3},c,l_{3}}_{x}\sigma^{h_{3}+1,\vec{p}_{3},c,l_{3}}_{y}\sigma^{h_{4},\vec{p}_{4},d,l_{4}}_{x}\sigma^{h_{4}+1,\vec{p}_{4},d,l_{4}}_{y}),\\
\end{eqnarray*}

\begin{eqnarray*}
\Omega_{4}&=&(-\sigma^{h_{3},\vec{p}_{3},c,l_{3}}_{y}\sigma^{h_{3}+1,\vec{p}_{3},c,l_{3}}_{x}\sigma^{h_{4},\vec{p}_{4},d,l_{4}}_{x}\sigma^{h_{4}+1,\vec{p}_{4},d,l_{4}}_{x}\\
&&-\sigma^{h_{3},\vec{p}_{3},c,l_{3}}_{y}\sigma^{h_{3}+1,\vec{p}_{3},c,l_{3}}_{x}\sigma^{h_{4},\vec{p}_{4},d,l_{4}}_{y}\sigma^{h_{4}+1,\vec{p}_{4},d,l_{4}}_{y}  \\
&& +\sigma^{h_{3},\vec{p}_{3},c,l_{3}}_{x}\sigma^{h_{3}+1,\vec{p}_{3},c,l_{3}}_{y}\sigma^{h_{4},\vec{p}_{4},d,l_{4}}_{x}\sigma^{h_{4}+1,\vec{p}_{4},d,l_{4}}_{x}\\
&&+\sigma^{h_{3},\vec{p}_{3},c,l_{3}}_{x}\sigma^{h_{3}+1,\vec{p}_{3},c,l_{3}}_{y}\sigma^{h_{4},\vec{p}_{4},d,l_{4}}_{y}\sigma^{h_{4}+1,\vec{p}_{4},d,l_{4}}_{y}\\
&&-\sigma^{h_{3},\vec{p}_{3},c,l_{3}}_{x}\sigma^{h_{3}+1,\vec{p}_{3},c,l_{3}}_{x}\sigma^{h_{4},\vec{p}_{4},d,l_{4}}_{y}\sigma^{h_{4}+1,\vec{p}_{4},d,l_{4}}_{x}\\
&&-\sigma^{h_{3},\vec{p}_{3},c,l_{3}}_{y}\sigma^{h_{3}+1,\vec{p}_{3},c,l_{3}}_{y}\sigma^{h_{4},\vec{p}_{4},d,l_{4}}_{y}\sigma^{h_{4}+1,\vec{p}_{4},d,l_{4}}_{x}\\
&&+\sigma^{h_{3},\vec{p}_{3},c,l_{3}}_{x}\sigma^{h_{3}+1,\vec{p}_{3},c,l_{3}}_{x}\sigma^{h_{4},\vec{p}_{4},d,l_{4}}_{x}\sigma^{h_{4}+1,\vec{p}_{4},d,l_{4}}_{y}\\
&&+\sigma^{h_{3},\vec{p}_{3},c,l_{3}}_{y}\sigma^{h_{3}+1,\vec{p}_{3},c,l_{3}}_{y}\sigma^{h_{4},\vec{p}_{4},d,l_{4}}_{x}\sigma^{h_{4}+1,\vec{p}_{4},d,l_{4}}_{y}).
\end{eqnarray*}
Note that the  repeated indices $a,b,c,d$, $\vec{p}_{1},\vec{p}_{2},\vec{p}_{3},\vec{p}_{4}$, $l_{1},l_{2},l_{3},l_{4}$ are not summed and the coefficients $W_{3,l_{1},l_{2},l_{3},l_{4}}, W_{4,l_{1},l_{2},l_{3},l_{4}}\in \mathbb{R}$ in Eqs.~(\ref{I21}, \ref{I22}) are defined by
\begin{eqnarray}
\epsilon_{\mu,l_{1}}\epsilon_{\nu,l_{2}}\epsilon^{\mu*}_{l_{3}}\epsilon^{\nu*}_{l_{4}} e^{-i \omega_{2}t}=W_{3,l_{1},l_{2},l_{3},l_{4}}+iW_{4,l_{1},l_{2},l_{3},l_{4}}.
\end{eqnarray}
{All of the coefficients appearing in quantum circuits are real valued number which maybe variant with time and qubit locations. These kinds of quantum circuits can be performed by quantum computer, for example, we can adjust the strength of laser beams to realize variable coefficients quantum circuits in trapped ion quantum computer.}

We need less than $65536(\mathcal{N}+1)^{4}$ CNOT operations to simulate $e^{-iI_{2}\Delta t}$ and 
$655360(N^{2}-1)^{5}\mathcal{V}^{3}(\mathcal{N}+1)^{4}$
CNOT operations to simulate $e^{-iH_{G4I}\Delta t}$ in a single time step. 

\section*{Appendix B: The Unitary Transformation $\mathcal{T}$}
\label{A2}
The unitary transformation $\mathcal{T}$ is 
\begin{eqnarray}
{\cal T} |q^{s_1}_{\vec{p}_{1},1} \bar{q}^{s_2}_{\vec{p}_{2},1}\rangle=\frac{1}{\sqrt{3}}\sum^{3}_{i=1} |q^{s_1}_{\vec{p}_{1},i} \bar{q}^{s_2}_{\vec{p}_{2},i}\rangle.
\end{eqnarray}
We label the states as
\begin{eqnarray}
|1\rangle=|q^{s_1}_{\vec{p}_{1},1} \bar{q}^{s_2}_{\vec{p}_{2},1}\rangle, 
|2\rangle=|q^{s_1}_{\vec{p}_{1},1} \bar{q}^{s_2}_{\vec{p}_{2},2}\rangle,
|3\rangle=|q^{s_1}_{\vec{p}_{1},1} \bar{q}^{s_2}_{\vec{p}_{2},3}\rangle,\nonumber\\
|4\rangle=|q^{s_1}_{\vec{p}_{1},2} \bar{q}^{s_2}_{\vec{p}_{2},1}\rangle, 
|5\rangle=|q^{s_1}_{\vec{p}_{1},2} \bar{q}^{s_2}_{\vec{p}_{2},2}\rangle,
|6\rangle=|q^{s_1}_{\vec{p}_{1},2} \bar{q}^{s_2}_{\vec{p}_{2},3}\rangle,\nonumber\\   \nonumber
|7\rangle=|q^{s_1}_{\vec{p}_{1},3} \bar{q}^{s_2}_{\vec{p}_{2},1}\rangle, 
|8\rangle=|q^{s_1}_{\vec{p}_{1},3} \bar{q}^{s_2}_{\vec{p}_{2},2}\rangle,
|9\rangle=|q^{s_1}_{\vec{p}_{1},3} \bar{q}^{s_2}_{\vec{p}_{2},3}\rangle,\end{eqnarray}
then $\mathcal{T}$ can be represented by the states as a $9\times 9$ matrix. We give an explicit representation as follow
\begin{widetext}
\begin{eqnarray}
\mathcal{T}=\frac{1}{\sqrt{3}}\left(
\begin{array}{ccccccccc}
1 & 0 &0&0& \frac{1}{2}  \left(-1+i\sqrt{3}\right) & 0&0&0&\frac{1}{2} \left(-i+\sqrt{3}\right) \\
0 & \sqrt{3} & 0 & 0&0&0&0&0&0 \\
 0 &0& \sqrt{3} & 0&0&0&0&0&0 \\
  0 & 0 & 0&\sqrt{3} & 0&0&0&0&0 \\
1 & 0 &0&0& -\frac{1}{2} \left(1+i\sqrt{3}\right) &0&0&0& -\frac{1}{2} \left(i+\sqrt{3}\right) \\
0 & 0&0 & 0&0 &\sqrt{3} &0&0&0\\
0 & 0&0 & 0&0 &0&\sqrt{3} &0&0\\
0 & 0&0 & 0&0 &0&0&\sqrt{3} &0\\
1 & 0&0&0 & 1 &0&0&0& i 
\end{array}
\right).
\end{eqnarray}

\end{widetext}

\end{document}